\documentclass[11pt]{article}

\usepackage[utf8]{inputenc}
\usepackage[T1]{fontenc}
\usepackage[english]{babel}
\usepackage{lmodern}
\usepackage[letterpaper,margin=1in]{geometry}
\usepackage{graphicx}
\usepackage{amsmath}
\usepackage{booktabs}
\usepackage{longtable}
\usepackage{array}
\usepackage{enumitem}
\usepackage{microtype}
\usepackage[round,authoryear]{natbib}
\usepackage[hidelinks]{hyperref}

\graphicspath{{figures/}}

\providecommand{\doi}[1]{doi: \href{https://doi.org/#1}{#1}}

\providecommand{\tightlist}{%
  \setlength{\itemsep}{0pt}\setlength{\parskip}{0pt}}

\title{\textbf{Applied and Filtered: An End-to-End Algorithmic Fairness Audit
of a Public Employment Agency}}
\author{Gemma Gald\'on-Clavell\\ \small Eticas.ai}
\date{}

\begin{document}
\maketitle

\begin{abstract}
Algorithmic fairness evaluation commonly assesses AI systems as bounded technical components, abstracting the organizational context in which they operate. We present, to our knowledge, the first independent end-to-end fairness audit of a semi-automated hiring system operated by Barcelona Activa, a public employment agency using the third-party TalentClue platform for candidate search and shortlisting. We analyze approximately 497,000 candidate-vacancy pipeline entries from September 2017 to September 2022.

We audit seven stages of the hiring pipeline, encompassing automated processing, human discretion, candidate data, and employer decisions. Aggregate outcomes across binary genders are statistically indistinguishable, yet this parity masks substantial disparities. Disaggregating the results reveals pronounced inequalities across salary levels, age groups, and gender identities. Women experience adverse impact in mid-salary shortlisting (DIR = 0.786, p \textless{} 0.001), alongside persistent salary disparities across 15 of 20 sectors and a compounded disadvantage for women aged 46--55 (DIR = 0.77). Non-binary candidates are shortlisted at less than one-third the rate of men (DIR = 0.295), although this estimate is limited by a small sample (N = 285). Age-related exclusion is also evident, with candidates aged 55 and over entirely absent from the pipeline despite comprising 15.6\% of Barcelona's labor force. These disparities are not static: over time, the gender gap in shortlisting narrows from 6.5 percentage points in 2017 to 1.3 percentage points in 2022.

The audit further reveals a vendor--deployer information asymmetry: Barcelona Activa lacks access to key information about TalentClue's matching logic and evaluation. Our findings show that fairness outcomes can arise from interactions among automated processing, human discretion, data quality, vendor opacity, and pipeline structure. We therefore build on prior calls for sociotechnical and end-to-end approaches to fairness evaluation, showing empirically why model-level assessment alone can be insufficient for understanding fairness in deployed systems.
\end{abstract}


\section{Introduction}

Recruitment is rarely a single decision point. It is a cumulative sequence of decisions through which candidates progress: vacancies are defined, requirements are translated into structured filters and free-text keywords, platforms surface candidate profiles, human reviewers exercise discretionary judgment, and shortlists are constructed for employers. Algorithmic tools increasingly support these processes across both public and private-sector hiring \citep{bogen2018,raghavan2020,ajunwa2023}, introducing potential sources of disparity at multiple stages. These disparities can compound as candidates progress through the pipeline: a candidate who is slightly less likely to appear in a platform's search results may also be less likely to receive human review and, consequently, substantially less likely to reach a shortlist. Such cumulative disadvantage is a characteristic feature of pipelined decision systems \citep{black2023,hellman2020}.

Despite the sequential and organizational nature of recruitment, algorithmic fairness evaluation commonly treats AI systems as bounded by technical components whose behavior can be assessed independently of the processes in which they are deployed. Fairness assessments typically focus on model inputs, outputs, or performance across demographic groups \citep{buolamwini2018,raji2019}. Regulatory approaches have similarly emphasized the automated tool. For example, New York City's Local Law 144 requires annual bias audits of automated employment decision tools (AEDTs), but does not require assessment of the broader organizational pipeline through which those tools influence hiring decisions. This creates a potential mismatch between what is evaluated and what determines outcomes for candidates: disparities may arise through interactions among automated processing, human discretion, data practices, organizational procedures, and downstream decisions rather than from the automated component alone.

This limitation is particularly consequential in high-risk domains such as employment, where automated systems are embedded within ongoing organizational processes. It also raises a temporal challenge. Candidate populations, labor-market conditions, analyst practices, and platform functionality can change over time, meaning that fairness properties may not remain stable between assessments. Effective evaluation therefore requires both broader scope and greater continuity: examining the full sociotechnical pipeline while monitoring relevant fairness indicators as the system operates in production. This builds on prior calls for sociotechnical approaches to algorithmic fairness and extends them to the empirical evaluation of deployed hiring systems.

These issues are increasingly relevant to regulatory practice. The EU AI Act classifies AI systems used in employment as high-risk under Article 6 and Annex III and establishes requirements including risk management, human oversight, and post-market monitoring (Article 72). Following the Omnibus Simplification Package agreed in May 2026, the compliance deadline for high-risk systems has been extended to December 2, 2027 \citep{eu2026}. In the United States, NYC Local Law 144, effective July 2023, provides an operational example of regulation centered on periodic bias audits. Early evidence concerning its implementation has raised questions about the effectiveness of this approach. \citet{wright2024}, for example, found that only 5\% of covered employers had publicly posted audit reports and that 96\% of published audits reported passing impact ratios.

We examine these issues through an independent end-to-end audit of a semi-automated hiring system operated by Barcelona Activa, the labor intermediation service of the Barcelona City Council. Barcelona Activa uses the third-party TalentClue platform to support candidate search and shortlisting for employers. We analyze approximately 497,000 candidate-vacancy pipeline entries spanning five years of operational data from September 2017 to September 2022. Rather than evaluating the platform as an isolated component, we examine seven stages of the recruitment pipeline encompassing automated processing, structured candidate data, human analyst discretion, and employer decisions. We assess disparities across stages and demographic intersections, benchmark pipeline participation against the local labor force, and examine changes in outcomes over time.

This paper makes four contributions:

\begin{enumerate}
\def\labelenumi{\arabic{enumi}.}
\item
  \textbf{End-to-end fairness audit of an operational hiring system.} We provide an independent audit of a semi-automated real-world hiring system using five years of operational decision data, extending fairness assessment beyond the automated component to the broader recruitment pipeline.
\item
  \textbf{Evidence of disparities masked by aggregate outcomes.} We show that aggregate gender parity can coexist with substantial disparities across pipeline stages, demographic intersections, salary levels, and sectors, demonstrating the limitations of relying on aggregate outcomes alone.
\item
  \textbf{Vendor--deployer information asymmetry as an audit finding.} We document how limited access to a third-party platform's matching logic and evaluation information constrains the deploying organization's ability to independently assess and govern the system.
\item
  \textbf{Evidence for continuous fairness evaluation.} We show that fairness outcomes change over time, demonstrating the limitations of point-in-time assessment and providing empirical support for continuous monitoring of deployed hiring systems.
\end{enumerate}

Together, these contributions build on prior calls for sociotechnical and end-to-end approaches to algorithmic fairness and provide empirical evidence for their application to deployed hiring systems.

\section{Related Work}

\subsection{From Component-Level Evaluation to Sociotechnical Auditing}

Algorithmic auditing has traditionally focused on evaluating discrete technical artifacts, establishing methodological foundations for documenting model behavior, dataset provenance, and demographic performance. Model Cards \citep{mitchell2019}, Datasheets for Datasets \citep{gebru2021}, and early algorithmic audits of commercial AI systems \citep{raji2019} established important practices for documenting and measuring disparities across demographic groups. These approaches remain valuable, but they provide limited visibility into how technical systems interact with the organizational processes in which they are deployed. A substantial literature has subsequently challenged component-level approaches to fairness evaluation. \citet{selbst2019} identified five abstraction traps that arise when fairness is conceptualized as a property of isolated technical components rather than sociotechnical systems. \citet{raji2020} extended auditing across the AI development lifecycle, while \citet{radiyadixit2023} demonstrated through audits of police facial-recognition deployments that systems may satisfy technical criteria while failing to meet broader sociotechnical requirements. \citet{lam2024} similarly developed an assurance-oriented auditing framework that emphasizes organizational context and accountability.

Recent surveys and frameworks suggest that this gap remains substantial. \citet{rismani2025}, analyzing 791 AI evaluation measures, found that existing evaluation predominantly focuses on models and outputs. \citet{weidinger2023} distinguish capability, human-interaction, and systemic-impact layers of AI evaluation and argue that existing practice concentrates primarily on capability-level assessment. \citet{birhane2024}, surveying 341 AI audits, further identify audit scope, organizational context, and stakeholder access as recurring limitations in the existing audit ecosystem. Other work has examined the conditions under which independent audits can be conducted. \citet{casper2024} argue that black-box access alone is insufficient for rigorous audits and that white- and ` outside-the-box' access to development, deployment, and organizational context are needed. Our audit illustrates the value of the latter, drawing on longitudinal operational data spanning the full recruitment pipeline. Finally, \citet{costanzachock2022} show how auditors' institutional positions shape the scope and outcomes of algorithmic audits.

Our contribution extends this literature by operationalizing the sociotechnical paradigm in an applied employment audit using real operational data, demonstrating specifically what model-only evaluation misses and quantifying the magnitude of the gap.

\subsection{Evidence of Bias in AI-Assisted Employment Systems}

A substantial empirical literature documents disparities in algorithmic systems used in employment and labor markets. Early studies demonstrated that ostensibly neutral advertising and ranking systems can reproduce demographic disparities. \citet{datta2015} found that women were shown fewer advertisements for high-paying jobs, while \citet{lambrecht2019} found that women were underrepresented in the delivery of STEM career advertisements despite gender-neutral targeting. \citet{imana2021} further showed that platform-side optimization can reproduce occupational gender disparities independently of advertisers' targeting decisions. Research on employment search and ranking systems has similarly identified gender disparities in resume ranking \citep{chen2018} and worker search and ratings on online labor platforms \citep{hannak2017}.

Research examining hiring technologies has highlighted additional challenges. \citet{raghavan2020} reviewed the practices and public claims of 18 pre-employment assessment vendors and found substantial variation in validation practices and attention to adverse impact. \citet{sanchezmonedero2020} identified gaps between vendors' claims about addressing discrimination and the evidence provided to substantiate those claims. \citet{ajunwa2023} similarly identifies opacity, proprietary information, and limited transparency as recurring challenges in the use of AI in employment.

Together, the literature suggests that disparities in algorithmic employment systems can emerge through multiple mechanisms, including training data, feature selection, ranking objectives, system configuration, and downstream use \citep{bogen2018,black2024}. It also highlights a distinction between evaluating the technical behavior of a hiring tool and evaluating the outcomes produced by the broader hiring process in which it is embedded. Our study addresses this distinction directly by tracing candidate outcomes across multiple stages of an operational recruitment pipeline rather than evaluating a single algorithmic component.

\subsection{Public Sector AI and the Auditing Vacuum}

Public-sector deployments of AI in employment and welfare have generated substantial research and scrutiny, but comparatively few independent evaluations of deployed systems. The Netherlands' System Risk Indication (SyRI), used to identify potential welfare fraud, was found unlawful by the District Court of The Hague in 2020 on Article 8 ECHR grounds \citep{rachovitsa2022}. Austria's public employment service (AMS) has been examined in relation to an algorithm that classified unemployed individuals into three tiers using, among other variables, gender and care obligations \citep{allhutter2020}; subsequent work has evaluated aspects of the system using empirical data \citep{achterhold2025}. Research on Germany's public employment service has similarly examined algorithmic decision-making using administrative data, although the evaluated models were not deployed \citep{kern2021}.

Despite this, there remains limited empirical evidence on how deployed public-sector AI systems actually behave within the organizational processes in which they are embedded. Our study addresses this gap through an independent end-to-end audit of an operational public employment system using five years of real decision data, examining disparities across the recruitment pipeline, demographic groups, and time.

\section{System Description}

\subsection{Barcelona Activa and TalentClue}

Barcelona Activa is the economic development agency of the Barcelona City Council, operating a labor intermediation service that connects registered job seekers with employer vacancies. Unlike fully automated recruitment systems, Barcelona Activa operates through a human-in-the-loop decision process: analysts interpret employer requirements, define search criteria, execute candidate searches on the third-party TalentClue platform, review returned profiles, and produce shortlists.

TalentClue is a proprietary recruitment platform that provides infrastructure for candidate search and filtering. It supports structured filters (categorical and numerical fields including age, gender, education level, experience, location), textual search across free-text profile descriptions, logical combination of search criteria, and ranking of results by configurable relevance criteria. The exact matching and ranking mechanisms used by TalentClue (whether exact match, fuzzy search, semantic similarity, or hybrid approaches), the weights assigned to different signals, and any internal fairness evaluation results are proprietary to the vendor and were not made available to Barcelona Activa or to the audit team.

\subsection{The Seven-Stage Pipeline}

Barcelona Activa's labor intermediation operates as a seven-stage pipeline, illustrated in Figure 1 and described below.

\begin{figure}[htbp]
  \centering
  \includegraphics[width=\linewidth]{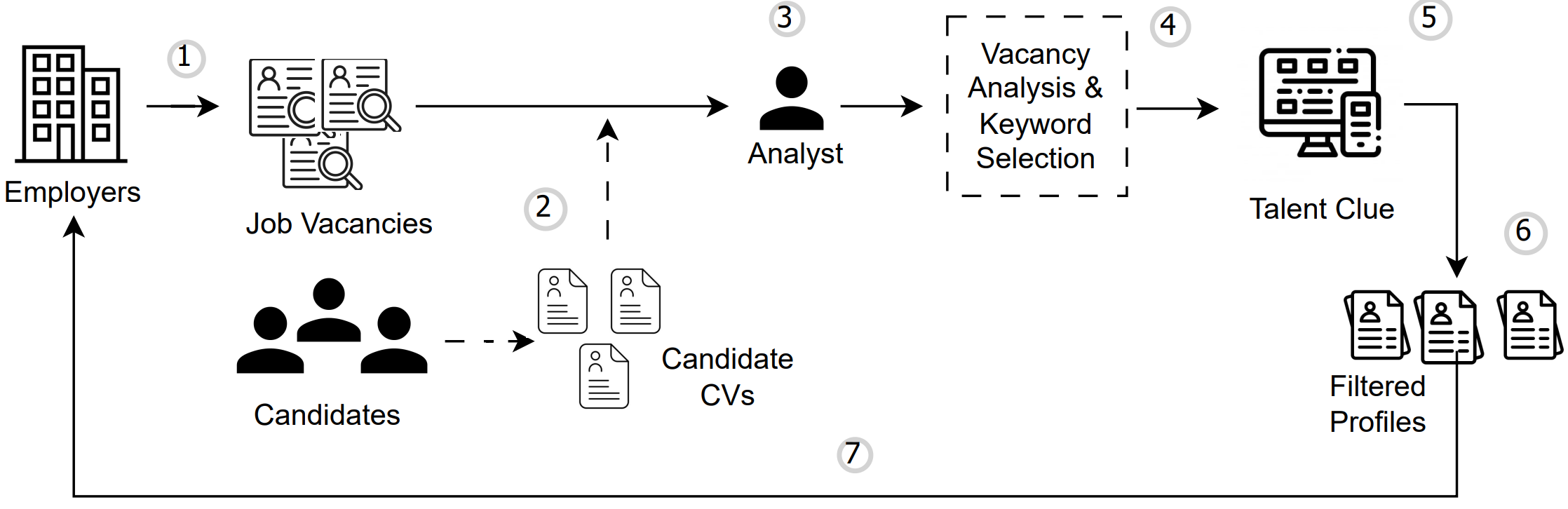}
  \caption{System architecture and data flow map. The dashed box indicates the scope of the audit and the role of Barcelona Activa in the recruitment pipeline.}
  \label{fig:1}
\end{figure}

\textbf{Stage 1: Vacancy receipt.} An employer submits a vacancy with job description, required qualifications or experience, desired skills, and working conditions (contract type, hours, salary range).

\textbf{Stage 2: Analyst assignment.} The vacancy is assigned to a Barcelona Activa analyst, who becomes the sole human operator mediating between employer requirements and platform capabilities. The analyst interprets requirements, executes searches, reviews results, and decides which candidates constitute an appropriate shortlist.

\textbf{Stage 3: Vacancy analysis and keyword selection.} The analyst translates qualitative requirements into structured filters and free-text keywords used to query the candidate database. No documentation, standardized terminology, or audit trail of keyword selection exists.

\textbf{Stage 4: Platform search.} The analyst queries TalentClue using structured filters and textual search, combining criteria with logical operators and ordering results by selected relevance criteria. The internal matching and ranking logic is not available to the deployer.

\textbf{Stage 5: Filtered profile results.} TalentClue returns a subset of candidate profiles matching defined criteria. The composition of the excluded pool is not recorded; Barcelona Activa has no way of assessing whether returned profiles are representative of the eligible candidate pool or whether certain groups are systematically less likely to appear.

\textbf{Stage 6: Final review and pre-selection.} The analyst manually reviews returned profiles and constructs the shortlist. This step operates without formal protocols, structured review criteria, or records of reasoning for inclusion or exclusion of specific candidates.

\textbf{Stage 7: Employer handoff.} The shortlist is delivered to the hiring employer, which conducts interviews and makes final decisions. Barcelona Activa does not systematically collect post-shortlist outcome data.

This architecture distributes decision authority across employers (Stage 1), platform logic (Stages 4 and 5), and analyst discretion (Stages 2, 3, and 6), with no feedback loop to detect discriminatory outcomes downstream (Stage 7). Each stage is a potential entry point for bias, and the absence of documentation at Stages 3 and 6 precludes retrospective inspection.

\subsection{Data}

The dataset covers September 2017 to September 2022 and contains approximately 497,000 pipeline entries representing candidate-vacancy pairings. Available attributes include declared sex (Men, Women, Non-binary/Other), age, country of origin, education level, sector, salary band, contract type (full-time/part-time), and pipeline status (registered, matched, shortlisted, hired, discarded). Critical missing data: 24.0\% of records lack gender information; 14.5\% lack country of origin. Candidate-level qualification, skills, and work experience data are not available in the dataset.

\section{Methodology}

\subsection{Scope and Risk Identification}

This audit applies Eticas' post-deployment algorithmic impact assessment methodology, developed and refined across Eticas' corpus of independent audits of deployed systems and formalized in the openly maintained Eticas AI Risk Taxonomy \citep{galdonclavell2026}. Prior to the quantitative analysis, a preliminary risk identification exercise mapped the system's risk profile against Eticas' AI risk taxonomy (arXiv:2607.02201; github.com/Eticas-AI/ai-risk-taxonomy), which classifies risks into twelve categories spanning Bias and Fairness, Privacy and Confidentiality, Reliability, Governance, Security and Misuse, Transparency and Explainability, Environmental Impact, Responsibility and Redress, Autonomy and Human Agency, Manipulation and Misinformation, Resilience, and Organizational Readiness. Per the conditions of the funding, the scope of this audit is restricted to Bias and Fairness. Privacy and Confidentiality and Governance were identified as potential risks warranting follow-up but are not assessed quantitatively in this work. Other risk categories (notably Reliability, Transparency, and Responsibility) are referenced where they intersect with fairness findings but are not separately evaluated. We note this scope restriction explicitly because the empirical findings and the audit's ability to identify them are conditioned on this scope.

Given that Barcelona Activa does not control TalentClue's model training or internal logic, in-processing checks are limited to what can be inferred from observable system outputs. Pre-processing analysis examines input data quality and representativeness; post-processing analysis examines operational decisions and temporal stability.

\subsection{Pre-processing: Representativeness Benchmarking}

We compare the demographic composition of the candidate pool against Barcelona's active labor force using the Spanish Labor Force Survey (Encuesta de Población Activa, EPA) for 2019--2021, disaggregated by gender, age, and region of origin, as compiled for the city by IDESCAT and the Barcelona City Council. Representativeness benchmarking establishes the baseline composition at the entry point of the system and tracks its evolution across pipeline stages, identifying points where specific groups are disproportionately filtered.

For AI systems with broad public-facing applications, reference populations are typically available from official statistics. The diagnostic question is simple: does the population the system serves reflect the population it claims to serve? Systematic mismatches indicate that the system has either failed to reach part of its intended population or has structurally excluded it.

\subsection{In-processing: Disparate Impact, Intersectional, and Stratified Analysis}

\textbf{Bias-risk mapping.} Before applying quantitative metrics, we map the seven-stage pipeline to identify decision points where selection filters, automated processes, or human judgments may introduce bias. This produces the risk classification reported in Table 1 (Section 5.1).

\textbf{Disparate Impact Ratio.} For each demographic group and pipeline transition, we compute the Disparate Impact Ratio:

\[DIR = \frac{\text{selection rate of protected group}}{\text{selection rate of reference group}}\]

The reference group is the highest-performing group for the outcome measured. A DIR below 0.80 is treated as the practitioner benchmark for adverse impact established by the EEOC Uniform Guidelines on Employee Selection Procedures \citeyearpar{eeoc1978} in the US. Following \citet{watkins2024}, we treat the 0.80 threshold as a benchmark, not a legal determination. We note that no equivalent numeric threshold exists in EU or Spanish law, though the legal concept of indirect discrimination on which DIR is based is well-established under EU equality directives.

\textbf{Statistical significance testing.} Fairness metrics are sensitive to sample size and stochasticity. We complement DIR with chi-square tests of independence, or Fisher's exact test for small samples, at \(\alpha = 0.05\). We report DIR and statistical significance jointly, alongside group sample sizes, recognizing that a large effect size in a small sample requires interpretation different from the same effect size in a large sample. Where DIR estimates fall close to the 0.80 threshold, we interpret with caution given sampling variability.

\textbf{Stratified conditional analysis.} Rather than relying solely on aggregate DIR, we stratify by sector, salary band, contract type, education level, and age group. Within-stratum DIR functions as a form of conditional analysis: it tests whether disparities persist after accounting for the covariate on which stratification is performed. This approach was selected over multivariate regression for three reasons: (a) interpretability for non-technical stakeholders is preserved; (b) stratified results are directly actionable at the operational level; (c) available covariates are limited (no qualification or experience data), constraining what a multivariate model could meaningfully condition on.

\textbf{Intersectional analysis.} We test combinations of protected attributes (sex × age, sex × education, sex × origin) for compounded disadvantage. Results are reported only where subgroup sizes permit reliable inference. Combinations falling below minimum support thresholds are not reported.

\subsection{Post-processing: Temporal Consistency}

Given the absence of post-shortlist outcome data from employers, post-processing analysis is restricted to signals available within Barcelona Activa's operational records. We compute shortlisting rates by gender for each year (2017--2022) and assess whether observed disparities are stable, narrowing, or widening. This temporal analysis tests the stationarity assumption underlying point-in-time auditing and informs the identification of monitoring indicators for ongoing fairness oversight.

\section{Results}

We report findings in order of analytical robustness. Findings supported by large samples (N \textgreater{} 100,000), within-group stratification, and statistical significance at p \textless{} 0.001 are presented as robust. Findings based on smaller subgroups or where confounding explanations cannot be ruled out are explicitly flagged as indicative.

\subsection{Pipeline Risk Mapping}

Table 1 maps the primary fairness risk at each stage of the pipeline and the extent of Barcelona Activa's visibility over it. Where visibility is limited or absent, the platform cannot currently detect, investigate, or correct the patterns its process produces.

{\small
\begin{longtable}[]{@{}
  >{\raggedright\arraybackslash}p{0.16\columnwidth}
  >{\raggedright\arraybackslash}p{0.44\columnwidth}
  >{\raggedright\arraybackslash}p{0.32\columnwidth}@{}}
\caption{Fairness and bias risks across the seven-stage Barcelona Activa hiring pipeline.}\label{tab:pipeline}\\
\toprule\noalign{}
\begin{minipage}[b]{\linewidth}\raggedright
Stage
\end{minipage} & \begin{minipage}[b]{\linewidth}\raggedright
Primary Risk
\end{minipage} & \begin{minipage}[b]{\linewidth}\raggedright
Visibility Gap
\end{minipage} \\
\midrule\noalign{}
\endhead
\bottomrule\noalign{}
\endlastfoot
1. Receipt of vacancy & Employer-defined requirements may encode discrimination within the job description (disproportionate experience criteria, gendered titles, below-market salaries for part-time roles). & No documented protocol for reviewing vacancy content before entry into the system. \\
2. Assignment to analyst & All subsequent judgments concentrate in a single analyst without documented criteria, structured decision rules, or oversight. & No mechanism to review, compare, or audit analyst decisions across vacancies. \\
3. Keyword selection & Translation of requirements into search terms is inherently subjective. Neutral-seeming keywords can carry implicit demographic associations. & No record of keywords selected for any vacancy; no documentation or audit trail. \\
4. TalentClue filtering & Structured filters explicitly include sex and age as filterable fields. Matching and ranking logic is proprietary. Systems trained on historical hiring patterns reproduce historical biases by design. & Deployer does not have access to platform's internal logic, weights, or evaluation results. Filter use by analysts is not recorded. \\
5. Filtered profile results & TalentClue returns only matching profiles. Composition of the excluded pool is not recorded. & No visibility into the composition of the excluded pool. \\
6. Analyst pre-selection & Shortlist decisions made without documented criteria, structured review, or records of reasons for excluding specific individuals. & No audit trail for inclusion or exclusion decisions. No consistency mechanism across analysts. \\
7. Employer selection & Employer reviews, interviews, and decides without protocol from Barcelona Activa and without outcome data being returned. & No post-shortlist data. No feedback loop. No mechanism to identify discriminatory employer practices. \\
\end{longtable}
}

Affected groups identified through this mapping and tracked across the pipeline include women (particularly in mid/high-salary and full-time roles), non-binary candidates, younger applicants (16--25), older adults (55+, absent from the pipeline entirely), and non-Spanish origin candidates. Observed disparities can arise through three non-exclusive pathways: candidate pool composition, TalentClue's filtering logic, or analyst discretion. These pathways cannot be fully disentangled with the available data.

\subsection{Pre-processing: Representativeness}

\textbf{Binary gender matches the labor market; non-binary candidates are nearly absent.} The candidate pool composition (51.5\% women, 48.5\% men) tracks the Barcelona EPA benchmark (47.6\% women, 52.4\% men), with women slightly overrepresented at registration. The categories ``Non-binary'' and ``Other'' appear with extremely limited representation (N = 285 across the audited period). This may reflect low population base rates, registration form design limitations (non-response, form normalization), or candidate reluctance to disclose for fear of discrimination or disclosure.

\textbf{Age composition shows structural exclusion of older adults.} The age distribution of the candidate pool broadly tracks the EPA benchmark for ages 16--54, with the youngest cohort moderately overrepresented and the 25--29 group showing the strongest convergence to the benchmark at the hiring stage (+6.5 percentage points). The most consequential finding is the absence of adults aged 55 and over: this group constitutes 15.6\% of Barcelona's active labor force but is effectively absent from the platform across all stages of the pipeline. The structural nature of this absence (zero, not merely underrepresented) indicates either coverage failure at registration, systematic exclusion through search practices or employer requirements, or limitations of the dataset (age truncation or cleaning rules) that themselves constitute a finding about the system's auditability.

\begin{figure}[htbp]
  \centering
  \includegraphics[width=\linewidth]{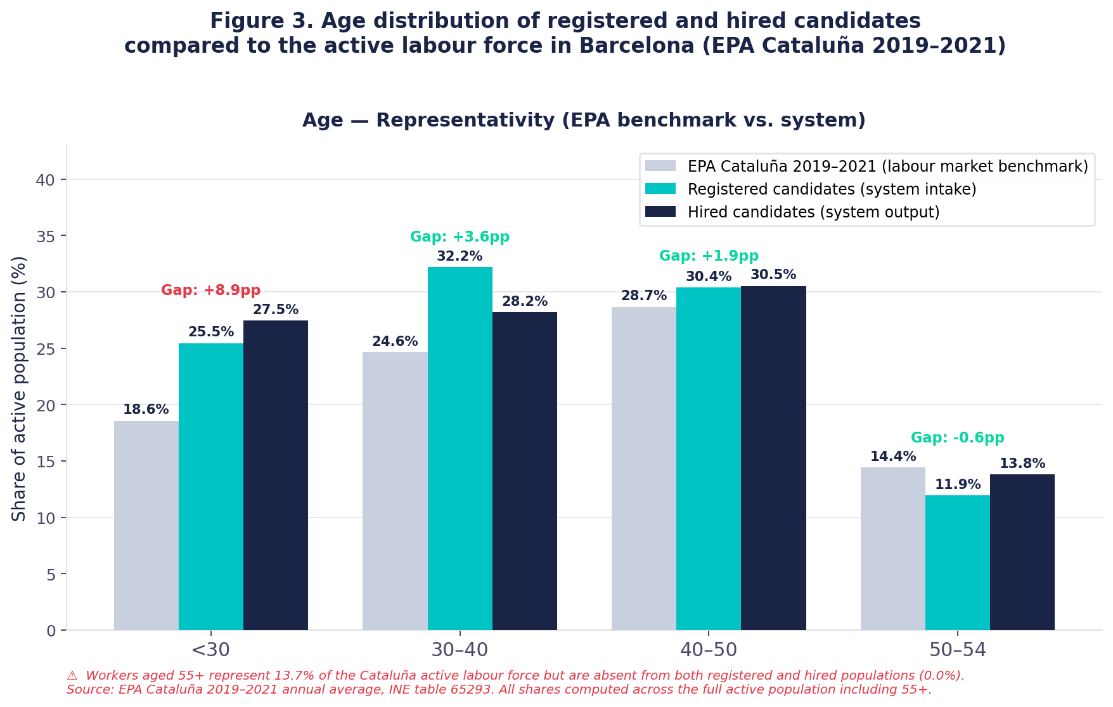}
  \caption{Age distribution of registered and hired candidates compared to the Barcelona active labor market. Adults aged 55+ are not shown as their representation in the platform is effectively zero, despite representing 15.6\% of the active labor force.}
  \label{fig:2}
\end{figure}

\textbf{Occupational composition shows sectoral clustering by gender.} Female candidates are substantially underrepresented relative to the Barcelona labor market in Real Estate and Architecture (-26.0 pp) and Engineering and Automotive (-16.1 pp). They are overrepresented in Logistics and Trading (+22.6 pp), Fashion and Apparel (+18.2 pp), and Human Resources (+15.2 pp). The corresponding pattern for male candidates is the mirror image. This sectoral clustering aligns with existing patterns of occupational segregation in the Spanish labor market but raises the question of whether the platform reproduces or attenuates these patterns. Because TalentClue's matching logic is not observable, the audit cannot determine whether observed disparities arise from candidate preferences, vacancy availability, or platform matching, but the persistence and magnitude of the clustering warrants concern.

\begin{figure}[htbp]
  \centering
  \includegraphics[width=\linewidth]{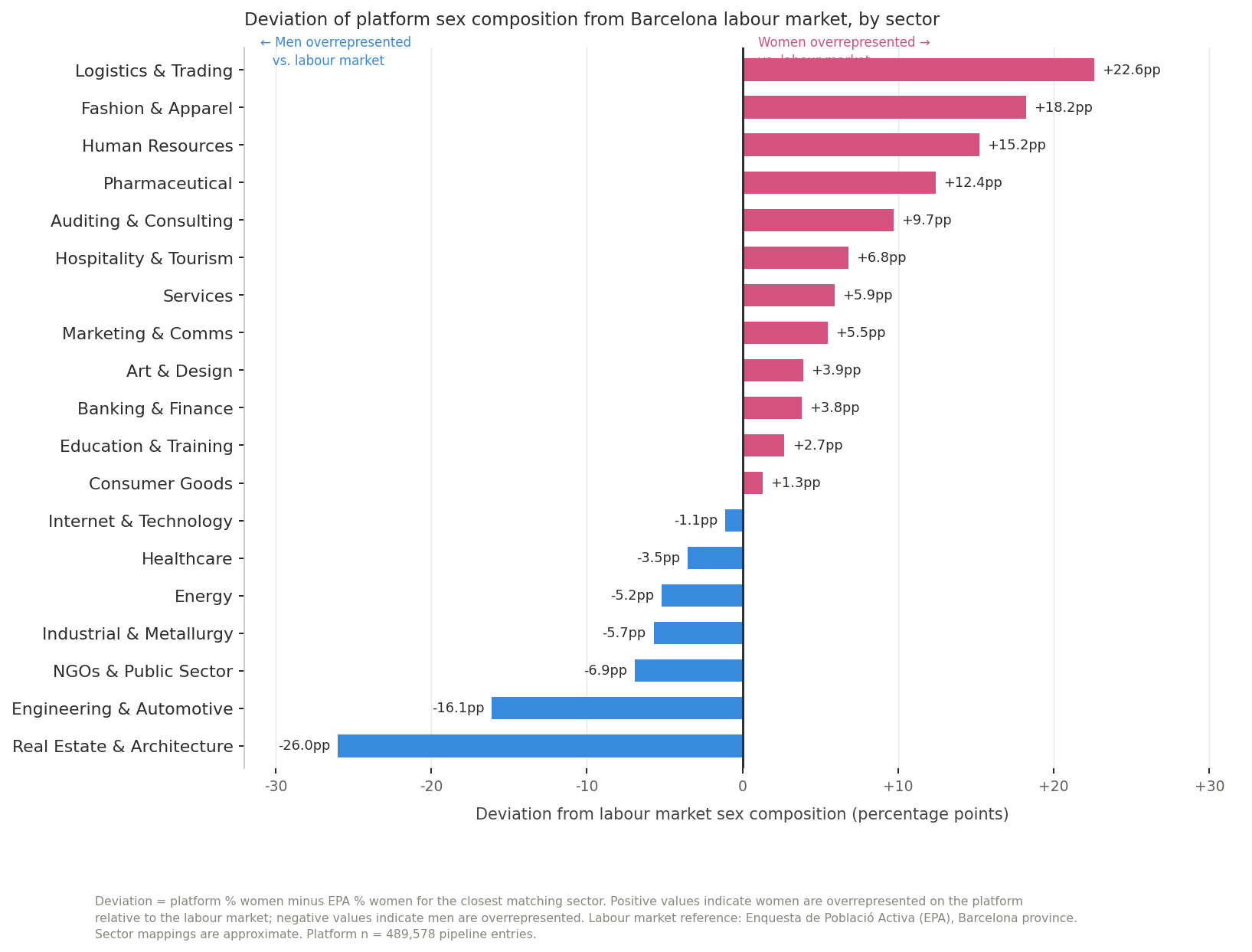}
  \caption{Deviation of platform sex composition from the Barcelona labor market by sector. Blue bars indicate men overrepresented vs.~labor market; pink bars indicate women overrepresented.}
  \label{fig:3}
\end{figure}

\subsection{In-processing: Gender Disparities in Shortlisting}

\textbf{Aggregate hiring outcomes are statistically equitable.} Women constitute 51.5\% of registered candidates and 49.8\% of hired candidates; the difference is not statistically significant. Under a model-output audit framework limited to aggregate selection rates, this system would be reported as fair. The remainder of this section demonstrates why such a report would be misleading.

\textbf{Sector-level shortlisting compounds occupational segregation.} Comparing women's share in matched candidate pools to their share in shortlists, women's representation drops statistically significantly in four sectors: Real Estate and Architecture (-12.4 pp, from 27.0\% to 14.6\%), Industrial and Metallurgy (-6.8 pp), Energy (-4.4 pp), and Logistics and Trading (-2.9 pp). Women's share increases significantly in Healthcare (+4.9 pp), where they already constitute a matched-pool majority of 70.7\%. The remaining fifteen sectors show no statistically significant shift. The interpretation is consequential: even within sectors where women are matched, shortlisting further reduces their representation in male-typed sectors. This is not attributable to who self-selects into the platform; it emerges from how matched candidates are filtered into shortlists.

\textbf{Within-sector salary gaps persist.} Women's profiles are matched to vacancies with an average minimum salary of €16,492, compared to €17,639 for men, a difference of €1,147 (p \textless{} 0.001). When stratified by sector, this gap persists in 15 of 20 sectors, statistically significant in 12, with the largest within-sector gaps in Internet and Technology (€2,012), Engineering and Automotive (€1,994), Art and Design (€1,727), and Real Estate and Architecture (€1,675). The persistence of within-sector salary gaps after sector stratification is the methodologically important finding: it demonstrates that sector composition alone does not explain the aggregate gap. Even when controlling for sector, women's profiles are systematically matched to lower-paying vacancies.

\begin{figure}[htbp]
  \centering
  \includegraphics[width=\linewidth]{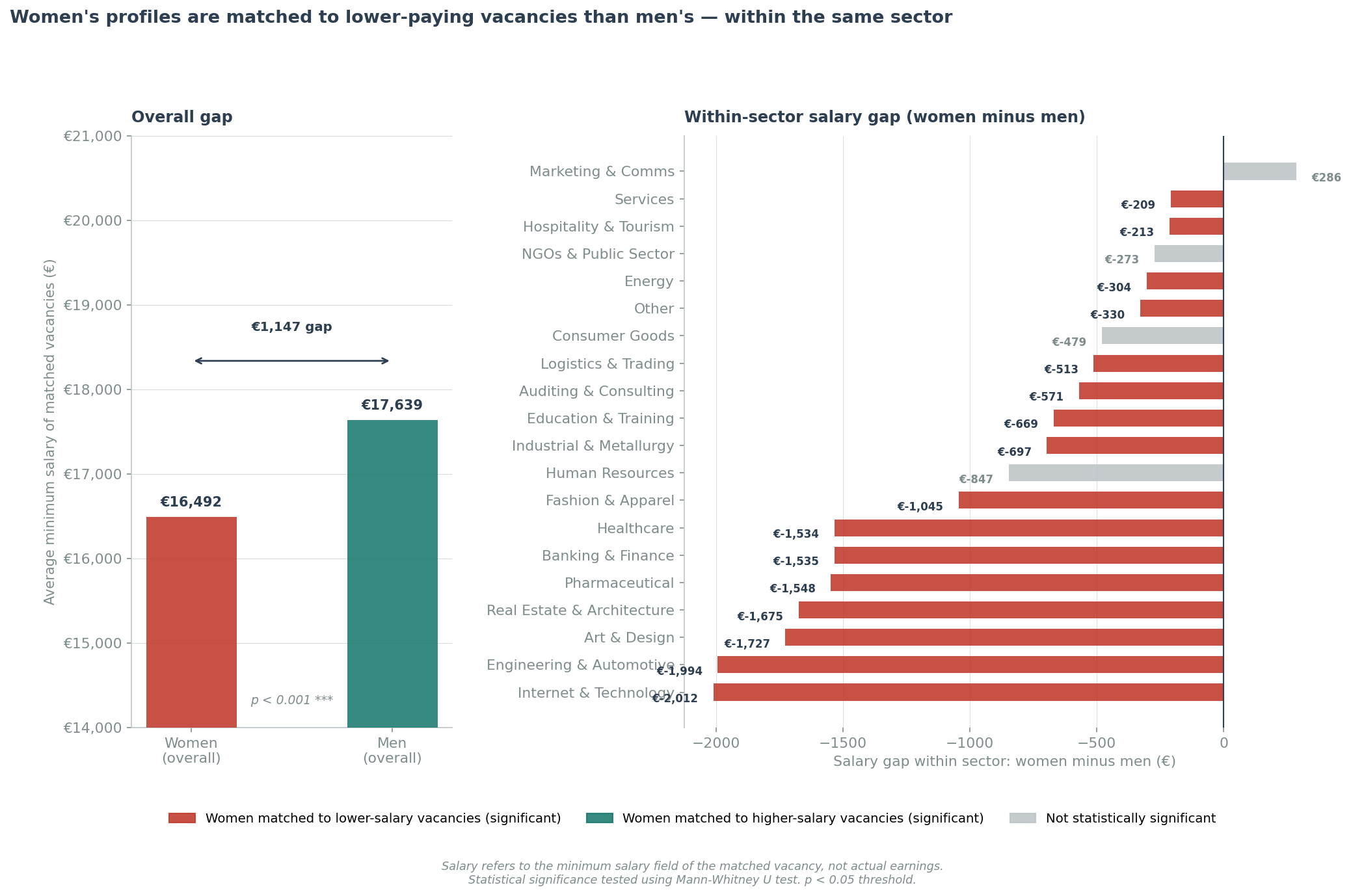}
  \caption{Comparison of salaries for men vs.~women (left) and within-sector salary gap (right). Red bars indicate women matched to lower-salary vacancies (statistically significant); grey bars are not statistically significant.}
  \label{fig:4}
\end{figure}

\textbf{Salary stratification reveals concentrated disadvantages at mid- and high-salary levels.} Shortlisting rates by gender across three salary bands show no adverse impact in low-salary roles (\textless€15,000; DIR = 1.010, p \textgreater{} 0.4) but adverse impact in mid-salary roles (€15,000--24,000; women 7.43\% vs.~men 9.45\%; DIR = 0.786, p \textless{} 0.001) and statistically significant though above-threshold disparity in high-salary roles (\textgreater€24,000; women 6.74\% vs.~men 8.12\%; DIR = 0.829, p \textless{} 0.001). The shortlisting gap is not uniform across salary levels; it concentrates where economic stakes are highest.

\begin{figure}[htbp]
  \centering
  \includegraphics[width=\linewidth]{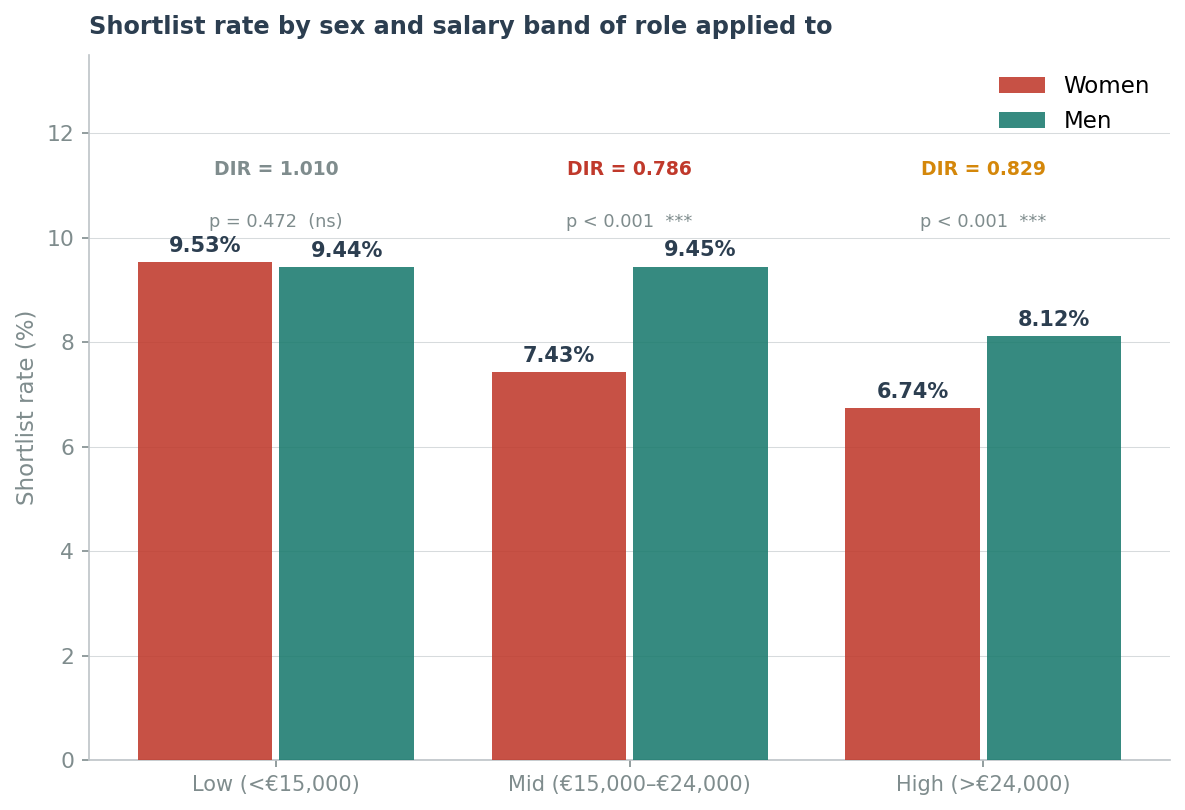}
  \caption{Shortlist rate for men and women across representative salary bands, with DIR scores and significance tests.}
  \label{fig:5}
\end{figure}

\textbf{Full-time roles show adverse impact.} Shortlisting rates for full-time positions show women at 9.00\% vs.~men at 11.87\% (DIR = 0.758, p \textless{} 0.001), below the 0.80 threshold. Part-time positions show women at 12.12\% vs.~men at 11.74\% (DIR = 1.032), with no adverse impact. The disparity is concentrated where job stability and earning potential are highest. Because candidates do not select which vacancies they are matched to (matching is analyst-driven), this disparity cannot be attributed to differential candidate preferences.

\begin{figure}[htbp]
  \centering
  \includegraphics[width=\linewidth]{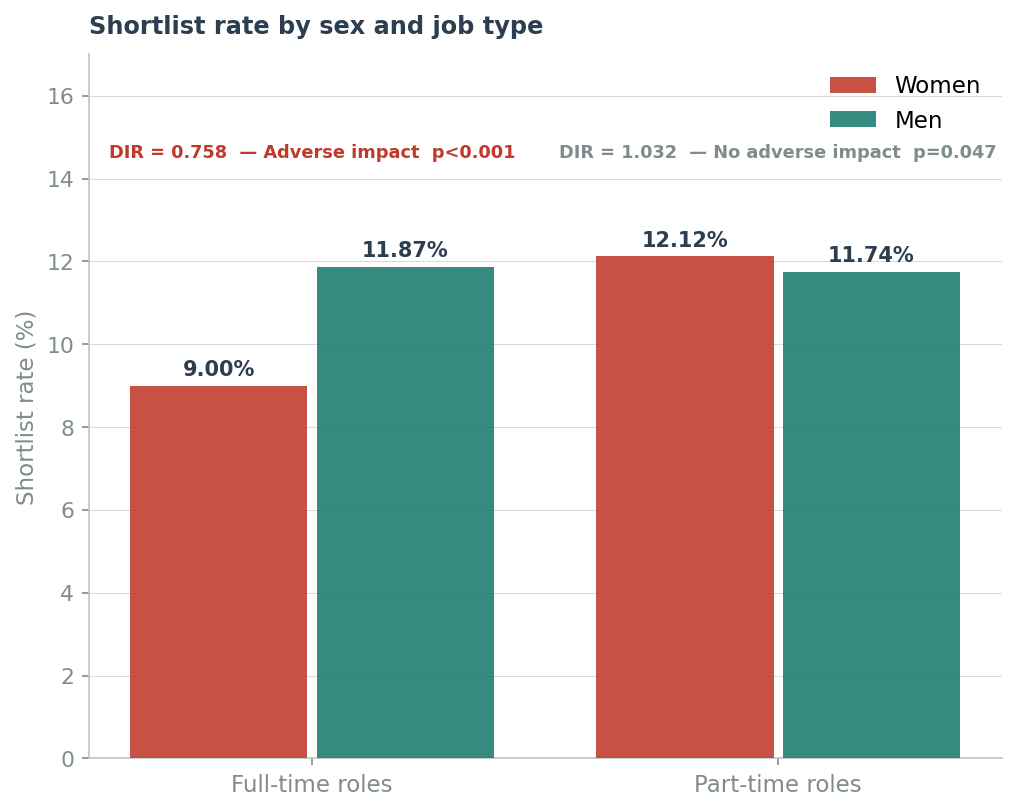}
  \caption{Shortlisting rate for men and women across part-time and full-time roles.}
  \label{fig:6}
\end{figure}

\subsection{Non-Binary Candidates}

Non-binary candidates are shortlisted at 3.51\% compared to 11.89\% for men, corresponding to DIR = 0.295 (Fisher's exact test, p \textless{} 0.001). The discard rate among non-binary candidates is 51.9\%, compared to 35.4\% for men and 36.9\% for women. Zero non-binary candidates were forwarded to employers across the audit period.

\begin{figure}[htbp]
  \centering
  \includegraphics[width=\linewidth]{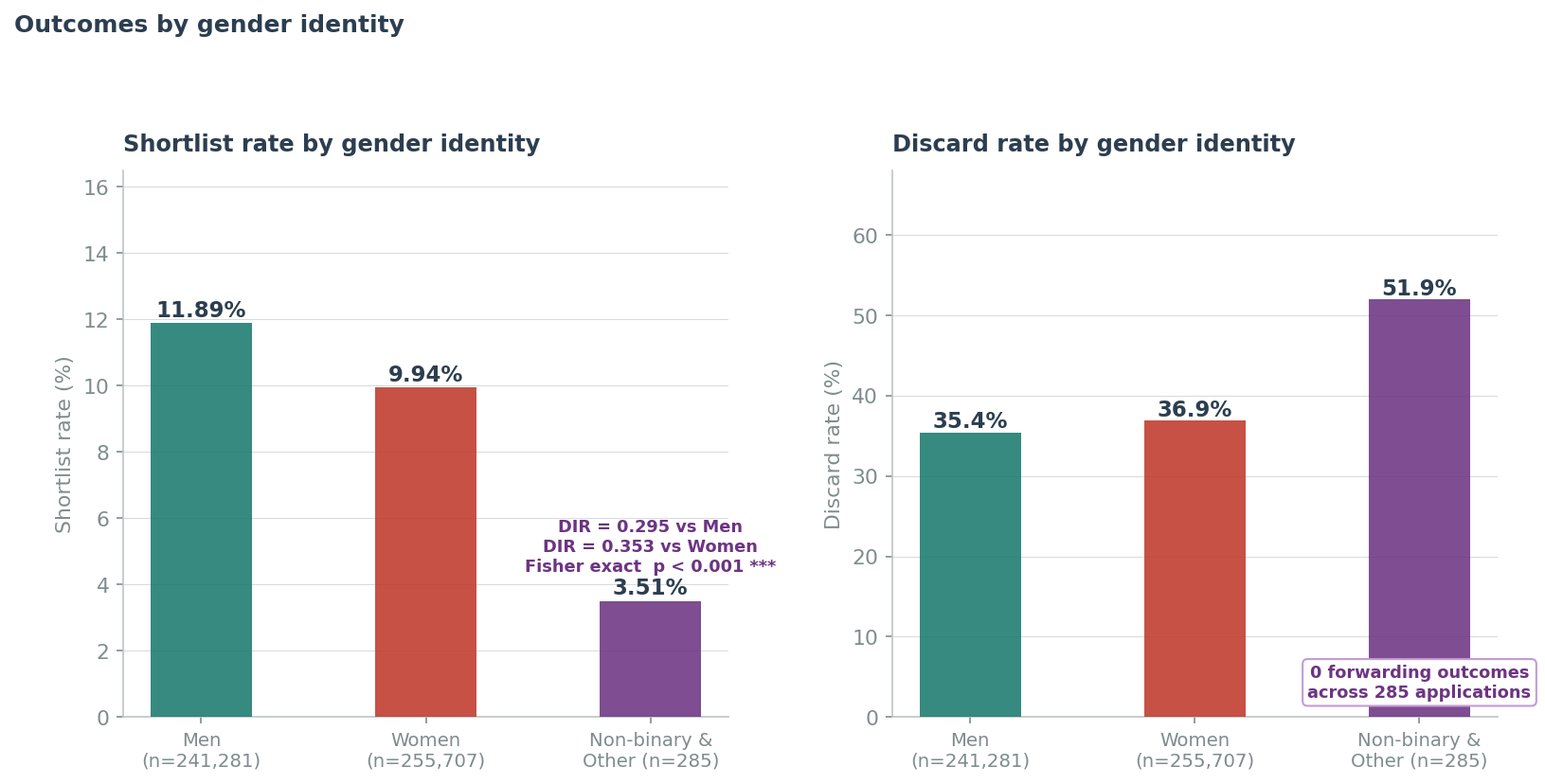}
  \caption{Shortlisting and discard rate for binary and non-binary identifying genders on the platform.}
  \label{fig:7}
\end{figure}

\textbf{We flag this finding as indicative rather than conclusive.} The sample (N = 285) is small relative to the binary gender groups (N \textgreater{} 240,000 each), and the population of non-binary candidates using Barcelona Activa's services may differ from the broader population in ways the available data cannot capture. The statistical test confirms the disparity is unlikely to arise from chance alone, but the precise magnitude of the disadvantage is uncertain. However, this result highlights that the platform is not producing positive outcomes for gender-diverse candidates, and that Barcelona Activa currently has no mechanisms to detect, monitor, or investigate such disparities. Nonetheless, more data is needed to invesitgate whether the disparity can be attributed to

\subsection{Age and Intersectional Effects}

\textbf{Aggregate hiring rates are similar across age groups, but the shortlisting stage shows variation.} Younger applicants (16--25) exhibit the highest discard rate (approximately 13\%) and the lowest shortlisting rate. Older applicants (46--55) show the highest shortlisting rate. Differences in placement (hiring) outcomes are smaller than differences in shortlisting outcomes, indicating that pipeline-internal disparities can attenuate before reaching the final outcome.

\textbf{Older women face compounded disadvantage.} Intersectional analysis of age × sex reveals that the largest gender gap occurs in the 46--55 age group: men are shortlisted at 13.3\%, women at 10.4\% (DIR = 0.77, below the 0.80 threshold). The gender gap is smallest among the youngest cohort (DIR = 0.94 for ages 16--25) and widens with age (DIR = 0.87 for 26--35; DIR = 0.93 for 36--45). The directional consistency of the gap across all age groups indicates a systematic pattern, with magnitude increasing among older applicants.

\begin{figure}[htbp]
  \centering
  \includegraphics[width=\linewidth]{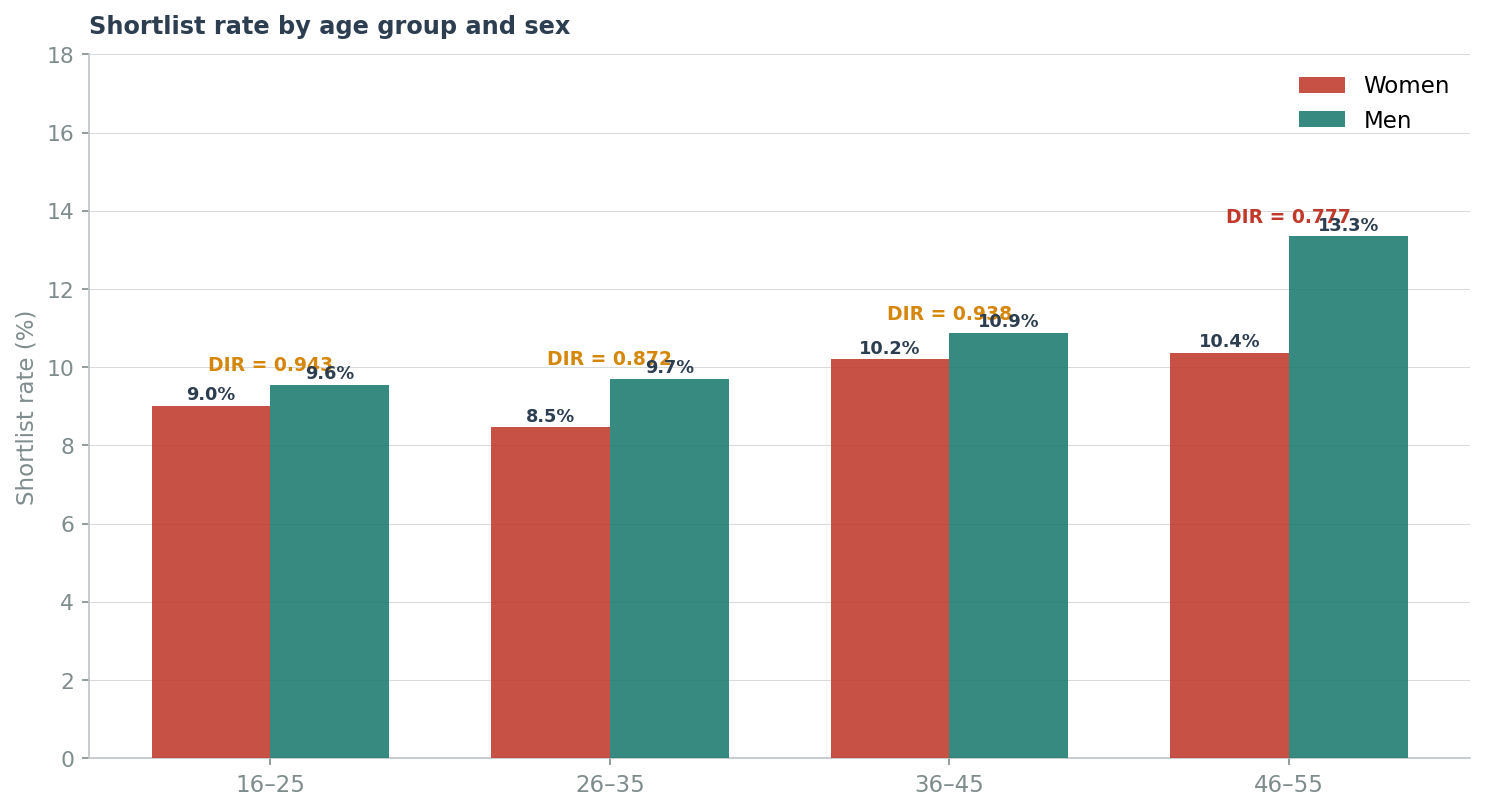}
  \caption{Shortlist rate for men and women across different age categories.}
  \label{fig:8}
\end{figure}

\subsection{Education}

Higher-educated candidates are shortlisted at rates 40--42\% lower than high school graduates (DIRs 0.576--0.600 for Graduate, Higher Level Training Cycle, Senior Level Vocational, Postgraduate, Training Not Completed, Degree, and Bachelor categories). This ``education paradox'' may reflect employer requirements for entry-level vacancies, candidate-vacancy mismatch, or filtering that penalizes overqualification.

\begin{figure}[htbp]
  \centering
  \includegraphics[width=\linewidth]{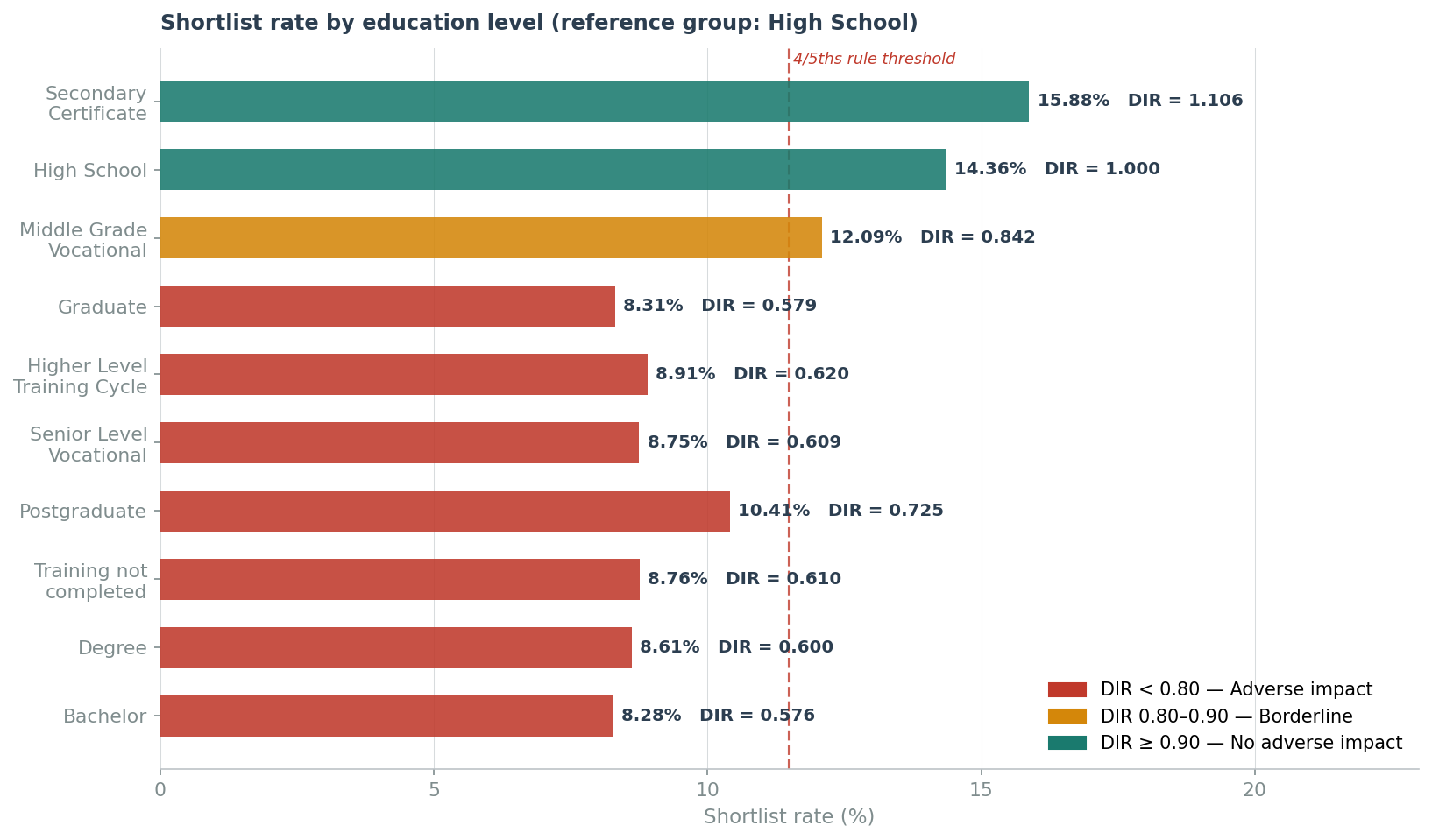}
  \caption{Shortlisting rates for candidates with different educational backgrounds. Red line denotes 4/5 Disparate Impact rule threshold.}
  \label{fig:9}
\end{figure}

However, because higher education categories skew female (58--62\% women for Degree, Graduate, and Higher-Level Training Cycle, versus 37\% for Secondary Certificate), either mechanism disproportionately affects women` s shortlisting outcomes, independent of any gender-specific process in the system. The pipeline appears to penalize overqualification in a way that disproportionately affects women, although causal attribution would require qualification data not available in the audit.

\begin{figure}[htbp]
  \centering
  \includegraphics[width=\linewidth]{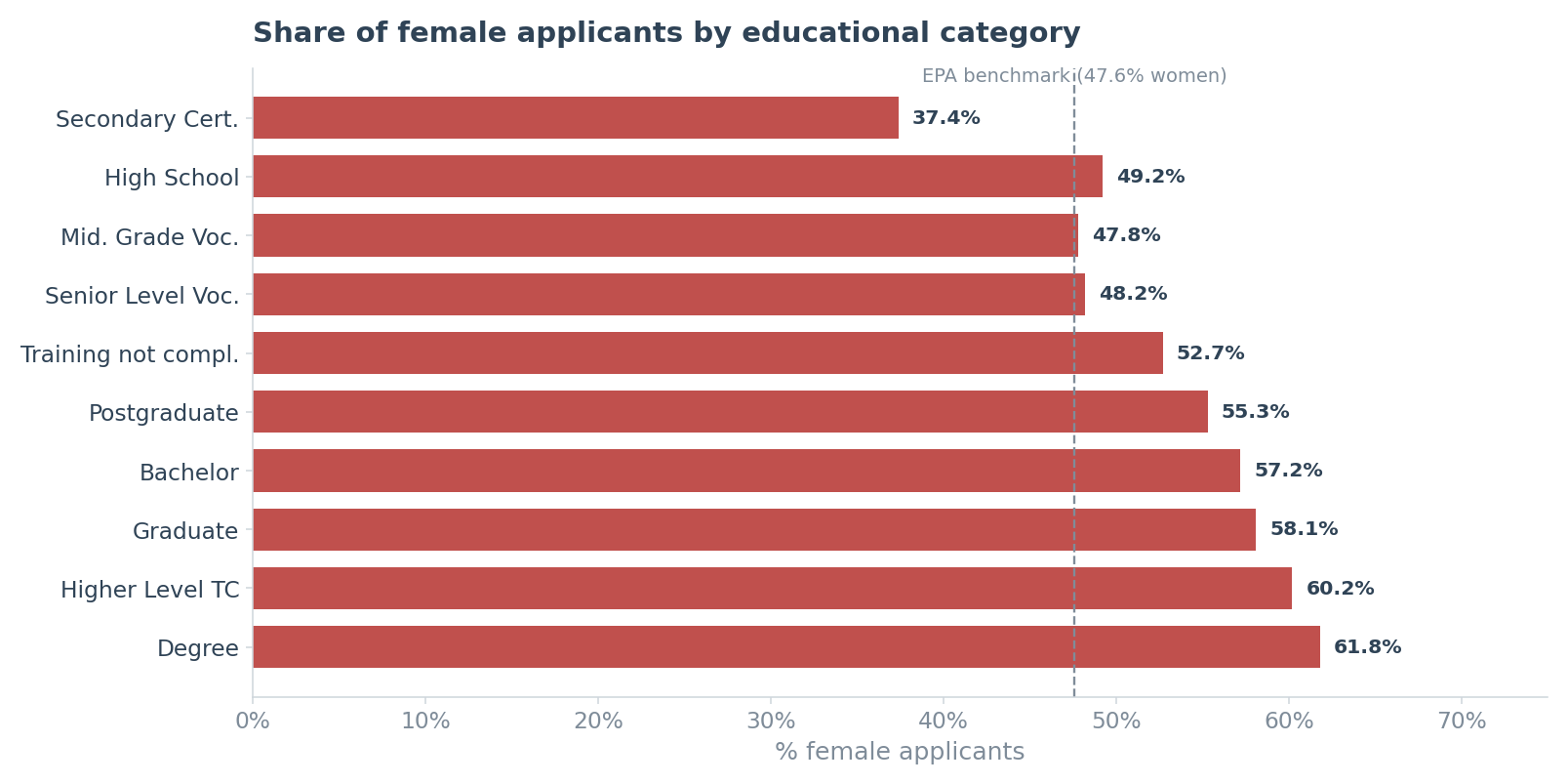}
  \caption{Percentage of female applicants in each educational category.}
  \label{fig:10}
\end{figure}

\subsection{Country of Origin}

Spanish candidates are shortlisted at 10.27\%, compared to 5.80\% for EU candidates (DIR = 0.565) and 6.48\% for non-EU candidates (DIR = 0.631). Both ratios fall below the 0.80 threshold. The disparity widens at the forwarding stage: EU and non-EU candidates are forwarded to employers at 0.200\% and 0.077\% respectively, against 0.518\% for Spanish candidates (DIR = 0.386 and 0.149; Figure 11, right panel).

\begin{figure}[htbp]
  \centering
  \includegraphics[width=\linewidth]{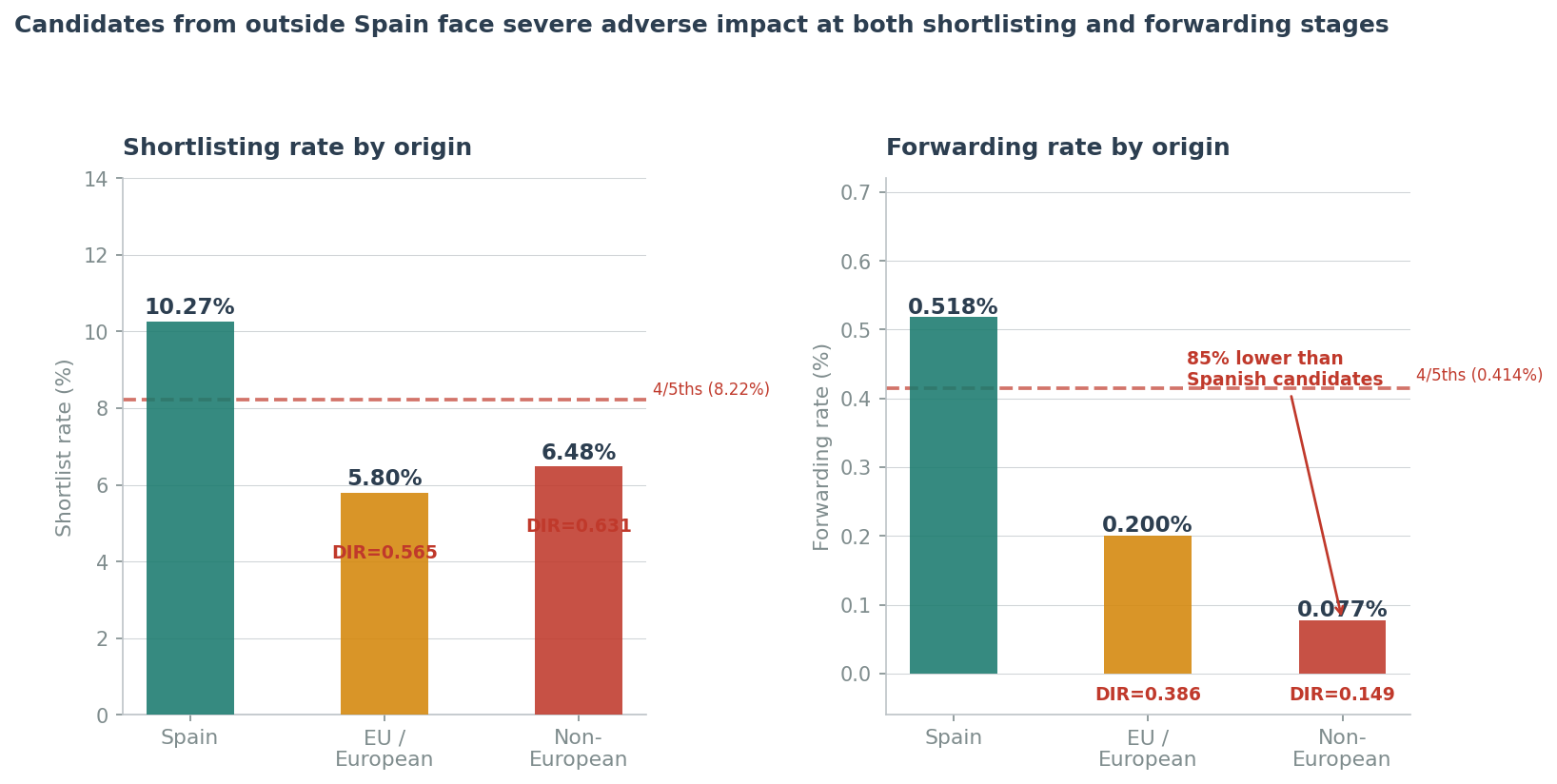}
  \caption{Shortlisting (left) and forwarding (right) rates by origin (Spanish, EU, non-EU candidates).}
  \label{fig:11}
\end{figure}

\textbf{We flag this finding as indicative.} Limitations include 14.5\% missing origin data, small non-EU subgroup (N = 242), and the absence of controls for language proficiency, credential recognition, or other factors plausibly correlated with origin. The pattern raises indirect discrimination risk under EU and Spanish equality frameworks, but further analysis with more complete data is required before drawing firm conclusions.

\subsection{Temporal Dynamics}

The shortlisting gap between men and women narrowed from approximately 6.5 percentage points in 2017 to less than 1.3 percentage points in 2022. This convergence, taken alone, would suggest improving fairness over time. However, it coincides with a decline in overall shortlisting rates for both groups, from approximately 18\% in 2017 to approximately 9\% in 2022. A plausible explanation is increased application volume relative to vacancies, producing higher competition and lower shortlisting rates across both groups. Whether the convergence reflects substantive fairness improvement or selection effects in a tightening pipeline cannot be determined from aggregated data.

\begin{figure}[htbp]
  \centering
  \includegraphics[width=\linewidth]{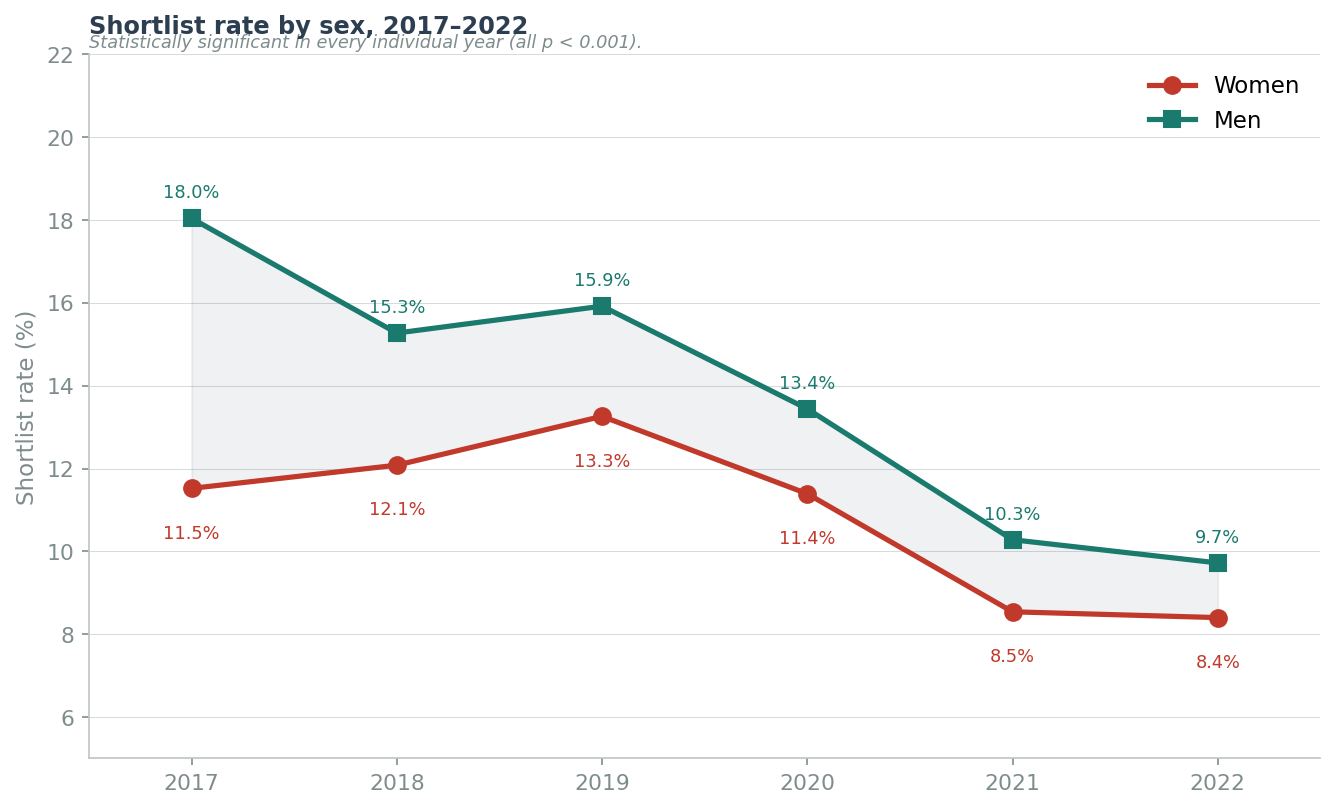}
  \caption{Change in shortlisting rate for men vs.~women, 2017--2022. The gap narrows over time, but overall shortlisting rates decline for both groups.}
  \label{fig:12}
\end{figure}

This finding has a structural implication for the auditing field: \textbf{any point-in-time audit produces a snapshot of a moving target.} An audit conducted in 2017 would have reported a large gender gap; the same audit in 2022 would report a small one. Neither snapshot captures the trajectory, and neither can determine whether the convergence will continue, stabilize, or reverse. Periodic auditing, however rigorous, is a necessary but insufficient accountability mechanism. We develop the implications of this point in Section 6.4.

\section{Discussion}

\subsection{What Model-Only Audits Miss: A Direct Comparison}

The contrast between Section 5.3 and the rest of Section 5 is, in our view, the central empirical contribution of this work. A model-output audit of Barcelona Activa's system would examine aggregate selection rates by gender across the registered-to-hired transition. The result (women 51.5\% of registered, 49.8\% of hired) shows no statistically significant disparity. A compliance audit limited to this scope, such as those mandated under NYC Local Law 144, would conclude that the system is non-discriminatory.

The pipeline audit reveals five disparities that the aggregate audit misses:

\begin{enumerate}
\def\labelenumi{\arabic{enumi}.}
\tightlist
\item
  \textbf{Mid-salary shortlisting:} DIR = 0.786, p \textless{} 0.001 (Section 5.3).
\item
  \textbf{Within-sector salary gaps:} persistent in 15 of 20 sectors (Section 5.3).
\item
  \textbf{Full-time shortlisting:} DIR = 0.758, p \textless{} 0.001 (Section 5.3).
\item
  \textbf{Compounded disadvantage at age × gender intersection:} DIR = 0.77 for women 46--55 (Section 5.5).
\item
  \textbf{Structural absence of adults 55+:} zero representation despite 15.6\% labor force share (Section 5.2).
\end{enumerate}

None of these findings appear in the aggregate hiring metric. All arise from the interaction of components across the pipeline: vacancy composition, analyst keyword choices, platform filtering, manual shortlisting, and employer-side selection. We argue that this gap between aggregate and pipeline findings is the rule, not the exception, in deploying AI-assisted decision systems.

The recent empirical literature on Local Law 144 implementation supports this concern. \citet{wright2024} reported that among 391 covered employers, only 5\% posted audit reports, and of those, 96\% of posted audits reported impact ratios above 0.80. \citet{gerchick2025} analyze the 116 publicly available audits and conclude that they constitute ``incomplete evaluations of algorithmic bias.'' Therefore, the compliance regime does not detect the disparities the regulation was designed to prevent.

\subsection{The Case for Continuous Evaluation in Production}

The temporal results (Section 5.8) show that fairness properties in this system are non-stationary. The gender gap in shortlisting narrowed from 6.5 to 1.3 percentage points over five years, although this change coincided with a decline in overall shortlisting rates. Thus, the system observed in 2022 cannot be assumed to have the same fairness properties as the system observed in 2017, even though no model retraining or deliberate intervention is documented.

This finding has implications for how fairness evaluation is operationalized. Existing regulatory frameworks, including the EU AI Act, NYC Local Law 144, Illinois HB 3773, and California and Colorado automated-decision regulations, generally rely on pre-deployment assessments, periodic audits, or assessments triggered by particular events. Academic auditing practices similarly tend to evaluate systems at discrete points in time. Such approaches provide important accountability mechanisms, but they offer limited visibility into changes between assessments. Operational systems can change as candidate populations shift, analyst practices evolve, platform functionality is updated, and labor-market conditions fluctuate. Consequently, fairness cannot be treated solely as a static property established through periodic assessment.

We therefore synthesize these strands into a continuous evaluation layer in production as a complementary accountability layer alongside pre-deployment assessment, periodic independent auditing, and incident-based investigation. Continuous evaluation differs from periodic auditing in three respects. First, it establishes an ongoing operational responsibility for the deployer while allowing monitoring infrastructure and analysis to be supported by independent evaluation providers. Second, it is prospective, enabling emerging disparities or changes in system behavior to be detected rather than documented only retrospectively. Third, it generates longitudinal evidence that can support subsequent independent audits and regulatory review. In this sense, continuous evaluation is not an alternative to formal auditing but an infrastructure layer that makes accountability over the operational lifetime of a system possible. This approach also has a regulatory basis. Article 72 of the EU AI Act requires providers of high-risk AI systems to establish post-market monitoring plans that systematically collect, document, and analyze performance data throughout the system` s lifetime. However, much of the data needed to evaluate employment outcomes---including application, filtering, and shortlisting patterns---is generated within the deployer` s operational environment.

\subsection{Limitations and Causal Interpretation}

The interpretation of our findings is subject to several limitations. First, the available data constrain the analyses that can be performed. The absence of qualification, skills, and experience variables prevents conditional parity analyses. In addition, 24\% of records lack gender information and 14.5\% lack origin information. The absence of analyst keyword data prevents reviewer-level analysis, and post-shortlist employer decisions are not captured. Second, the observational design of the audit limits causal interpretation. Although we identify disparities across multiple stages of the recruitment pipeline, the available data do not allow us to disentangle the relative contributions of candidate pool composition, platform functionality, and human decision-making. Finally, the scope of the audit was limited to fairness and bias outcomes; other risk dimensions, including privacy, governance, and reliability, were not evaluated.

Accordingly, the disparities identified in this study should be interpreted as evidence of differential outcomes and potential fairness risks rather than as causal effects attributable to any individual component of the system. Differences in shortlisting rates, salary matching, and candidate progression may reflect the interaction of multiple mechanisms, including applicant characteristics, platform processing, and analyst discretion. The available evidence does not allow these mechanisms to be independently identified or their relative contributions to be quantified.

\section{Recommendations}

Based on the audit findings, we identify operational recommendations for organizations deploying semi-automated decision systems and policy recommendations for regulators and public agencies responsible for governing their use.

\subsection{Operational Recommendations}

\textbf{Establish fairness-oriented data infrastructure.} Organizations should develop the data collection, logging, and review processes necessary for ongoing fairness assessment. This includes improving the completeness of demographic and candidate profile information, documenting missing data patterns, and ensuring access to qualification, skills, and experience variables required for meaningful analysis.

\textbf{Standardize human decision processes.} Organizations should establish auditable protocols for analyst interactions with recruitment platforms, including documentation of search terms, standardized occupational terminology, and training on non-discrimination obligations. Recording human interventions is necessary to evaluate how automated tools interact with discretionary decision-making.

\textbf{Strengthen vacancy and vendor governance.} Analysts should be empowered to identify potentially exclusionary job requirements before candidate searches are conducted. In parallel, procurement processes for third-party systems should require access to documentation, evaluation results, and operational data necessary for independent assessment.

\textbf{Implement continuous fairness monitoring.} Organizations should establish ongoing monitoring of relevant fairness indicators across demographic groups and decision stages. Monitoring should include defined review procedures and escalation mechanisms when disparities exceed established thresholds. Continuous evaluation complements, rather than replaces, independent audits by providing visibility into changes in system behavior over time.

\subsection{Policy Recommendations}

\textbf{Auditability as a procurement requirement.} Public agencies deploying high-risk AI systems are already subject to emerging regulatory requirements (e.g., under the EU AI Act) that mandate varying levels of transparency, documentation, and oversight. However, procurement processes should operationalize these obligations by requiring, prior to deployment, access to sufficient documentation, operational data, and evaluation information to support independent assessment. Without clear procurement-level requirements, deployers may still lack the information necessary to effectively govern third-party systems.

\textbf{Regulate outcomes and accountability mechanisms rather than fixed evaluation procedures.} Regulatory requirements should ensure that high-risk AI systems are evaluated against meaningful outcomes within their deployment context, rather than relying solely on predefined audit procedures. Because relevant fairness considerations vary across domains, populations, and decision processes, organizations should be required to justify their evaluation methodology, chosen benchmarks, and monitoring approach. Such an approach would preserve regulatory flexibility while encouraging evaluations that better reflect the real-world operation of deployed systems.

\textbf{Strengthen implementation of post-market monitoring requirements.} Existing regulations, such as Article 72 of the EU AI Act, recognize the need for monitoring high-risk AI systems throughout their lifecycle. Effective implementation requires clear mechanisms for ensuring that relevant operational data and system performance information are available to the actors responsible for evaluating system behavior in deployment contexts. For third-party systems, this may require stronger coordination between providers and deployers to ensure that monitoring obligations can be meaningfully fulfilled.

\section{Conclusion}

This audit demonstrates that fairness in employment systems cannot be adequately assessed through evaluations of automated components alone. Our analysis of a semi-automated recruitment system operated by Barcelona Activa shows that aggregate parity can coexist with substantial disparities across decision stages, demographic groups, and employment sectors. These findings provide empirical evidence consistent with established work suggesting that fairness outcomes are shaped not only by system design and technical implementation, but also by human discretion, data practices, and the organizational processes surrounding system deployment.

More broadly, this work highlights a gap between existing approaches to algorithmic evaluation and the operational realities of deployed AI systems. Effective accountability requires extending evaluation beyond model outputs to the broader sociotechnical pipeline, ensuring that deployers have access to the information necessary to assess third-party systems, and establishing mechanisms for continuous monitoring of system behavior over time. The temporal variation observed in this audit further demonstrates that fairness cannot be treated as a static property established through periodic assessment alone.

In this work, we present the first independent end-to-end fairness audit of a semi-automated hiring system operated by a public employment agency using longitudinal operational data. By examining how disparities emerge across recruitment stages and evolve over time, this work provides empirical evidence for advancing algorithmic fairness evaluation toward more comprehensive, sociotechnical, and continuous forms of accountability.

\section*{Acknowledgments}
This work was conducted as part of the EU Horizon Europe research project FINDHR (Fairness and Intersectional Non-Discrimination in Human Recommendation), Grant Agreement No.~101070212. We thank Barcelona Activa and their team for cooperation throughout the audit process. We are grateful to Carlos Castillo (Universitat Pompeu Fabra) for enabling the collaboration with Barcelona Activa. We thank Ariane Aumaitre, Rubén González, and Luis González for their contributions to the initial scoping, preliminary data processing, and preparation foundational to the quantitative analysis, and Alexandra Magaard for editorial and research support on the current version.

\nocite{groves2024,solaiman2023,sweeney2013}

\small
\bibliographystyle{apalike}
\bibliography{references}

\end{document}